\documentclass[twocolumn,floatfix,prl]{revtex4}
\usepackage{psfrag}
\usepackage{graphicx}
\usepackage{dcolumn}
\usepackage{bm}
\usepackage{natbib}

\begin{document}

\title{
Toward a description of contact line motion at higher capillary numbers
     }

\author{
Jens Eggers
}

\affiliation{
School of Mathematics, 
University of Bristol, University Walk, \\
Bristol BS8 1TW, United Kingdom 
        }

\begin{abstract}
The surface of a liquid near a moving contact line is 
highly curved owing to diverging viscous forces. Thus,
microscopic physics must be invoked at the 
contact line and matched to the hydrodynamic solution 
farther away. This matching has already been done for a variety 
of models, but always assuming the limit of vanishing speed.
This excludes phenomena of the greatest current interest,
in particular the stability of contact lines. 
Here we extend perturbation theory to arbitrary
order and compute finite speed corrections to 
existing results. We also investigate the impact of 
the microscopic physics on the large-scale shape of the interface.
\end{abstract}

\maketitle
The moving contact line problem is a famous example
of  hydrodynamics failing to describe a macroscopic flow phenomenon. 
But it was only in 1971 that Huh and Scriven \cite{HS71} discovered 
that the viscous dissipation in the fluid wedge bordered by a 
solid and a fluid-gas interface is logarithmically infinite
if the standard no-slip boundary condition \cite{LL84} is applied
at the solid surface. Thus infinite force would be required to 
submerge a solid body, and a drop could never spread on a table.

This result is of course contradicted by observation, and 
physical effects that relieve the singularity have to be invoked 
near the contact line, which go beyond the standard description.
A great variety of possible mechanisms 
have been proposed, and indeed there is no reason to believe
that for different solid-fluid-gas systems always the same mechanism
is involved. However, a question rarely considered is whether 
the choice of different microscopic mechanisms would make a 
great difference when looked at macroscopically. 

In a recent paper \cite{ES03} we compared various microscopic 
models in the case of perfect wetting. We found that
the length scale that appears in the expression for the interface 
shape is {\it strongly} speed dependent, in a 
fashion that depends on the model. Here we are going to show that 
this dependence is much weaker in the case of partial wetting,
and differences only come in at higher order in an expansion in 
capillary number $Ca=U\eta/\gamma$, where $\eta$ is the viscosity 
of the fluid, and $\gamma$ the surface 
tension between liquid and gas. Finite capillary number corrections are of 
interest for various situations of ``forced'' wetting, in which $Ca$
is no longer asymptotically small, and previous 
theories for the dynamic interface angle break down \cite{K93}.

\begin{figure}
\includegraphics[width=1\hsize]{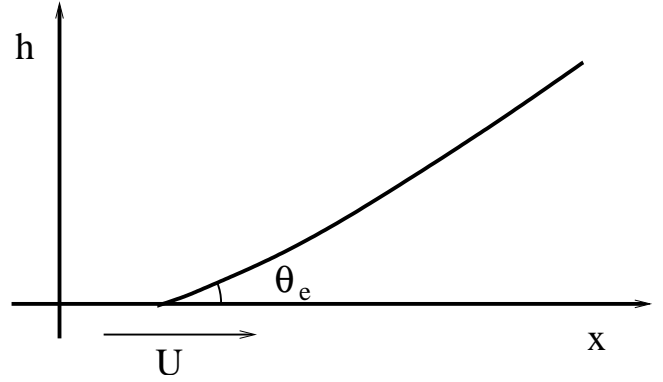}
\caption{\label{fig1} 
 A schematic of the interface near the contact line: In the frame of 
reference in which the contact line is stationary, the solid moves 
to the right in the ``wetting'' situation considered here. At the 
contact line, $h(0) = 0$, the slope of the interface is $\theta_e$.
    }
\end{figure}

For simplicity, we perform our 
calculations within the framework of lubrication theory, assuming 
a parabolic flow profile. This limits applications to the case of small 
contact angles, but without altering the essential structure of the 
problem. We consider the neighborhood of a contact line moving with speed
$U$ across a solid in a frame of reference in which the contact
line is fixed at the origin of the coordinate system (see Fig.\ref{fig1}). 
To relieve the corner singularity, we allow the fluid to slide across the 
solid surface, following the generalized Navier slip law 
\cite{N23,H83,H01}
\begin{equation}  
\label{Navier}  
u|_{y=0}-U = \lambda^{2-\alpha}h^{\alpha-1}\frac{\partial u}{\partial y}|_{y=0}
\end{equation}
at the plate, where $h(x)$ is the thickness of the fluid layer,
and $\lambda$ is taken as a constant rather than a speed dependent 
quantity. 
The case $\alpha=1$ corresponds to the usual Navier
slip, and (\ref{Navier}) is a simple generalization involving
only a single length scale $\lambda$. 
The resulting lubrication equation is \cite{H83}
\begin{equation}  
\label{lub}  
\frac{3 Ca}{h^2 + 3\lambda^{2-\alpha} h^{\alpha}} = -h'''.
\end{equation}  

The left-hand side corresponds to viscous forces, diverging 
as the contact line position $h(0)=0$ is approached, but weakened 
by the presence of slip. Viscous forces are balanced by surface 
tension forces on the right, resulting in a highly curved interface 
near the contact line. In comparison, other forces like gravity 
have been neglected. This restricts the 
validity of (\ref{lub}) to a distance from the contact line
below the capillary length $\ell_c=\sqrt{\gamma/(\rho g)}$.
We also assume that the angle at the contact line $h'(0)=\theta_e$
is constant, independent of speed. Hence it has to
coincide with the equilibrium contact angle, 
in order to give the right result at vanishing speed. 

Since we want to investigate the neighborhood of the contact
line, it is convenient to introduce the scaled variables
\begin{equation}  
\label{scal}  
h(x) = 3^{1/(2-\alpha)}\lambda 
H(\xi), \quad \xi=x\theta_e/[3^{1/(2-\alpha)}\lambda], 
\end{equation}  
which leads to 
\begin{equation}  
\label{sim}  
\frac{\delta}{H^2 + H^{\alpha}} = -H'''.
\end{equation}  

From the scaling (\ref{scal}) it is evident that the curvature
of the interface $h''(x)$ scales like $\lambda^{-1}$, where 
$\lambda$ is in the order of nanometers \cite{ES03}. Thus,
in order to match the local solution near the contact line 
to an outer profile with a curvature of order $1/\ell_c$, the curvature
$H''(\xi)$ has to {\it vanish} for large $\xi$. 
This means the boundary conditions for the solution of (\ref{sim}) are
\begin{equation}  
\label{bc}  
H(0) = 0, \quad H'(0)=1, \quad H''(\infty)=0.
\end{equation}  
The only parameter appearing in the problem is now the rescaled 
capillary number $\delta=3Ca/\theta_e^3$. 

For $\delta > 0$ equations (\ref{sim}) and (\ref{bc}) have a unique 
asymptotic solution, due to Voinov \cite{V76}, for which the 
slope behaves like $H'(\xi)=\left[3\delta\ln(\xi/\xi_0)\right]^{1/3}$ for
$\xi\gg 1$. This solution has vanishing curvature at infinity
and only contains a single free parameter $\xi_0$, to be determined by 
matching to the contact line. In the present paper, we are 
going to deal exclusively with this {\it wetting} situation.
If $\delta <0$, the mathematical structure of (\ref{sim}) changes
completely. This can be seen from considering the simpler equation 
$\delta/H^2 = -H'''$, valid for large $H$. Namely, it follows from 
an exact solution \cite{DW97} to this equation, that for $\delta < 0$ 
all solutions have strictly positive curvature at infinity. 
The consequences of this observation for the {\it stability} of 
contact lines are explored in another paper \cite{E03}.

The mathematical problem
to be tackled in this letter consists in computing $\xi_0$ as 
a function of $\delta$. As was done in previous works 
\cite{H83,C86,GHL90,H92}, 
we proceed by expanding around the trivial solution at {\it zero} speed
$\delta=0$. In this case, the solution of (\ref{sim}) and (\ref{bc}) is 
evidently given by $H(\xi)=\xi$. Hence the perturbation expansion 
we seek looks like 
\begin{equation}  
\label{pert}  
H'(\xi) = 1 + \delta H_1'(\xi) + \delta^2 H_2'(\xi) + \dots
\end{equation}  
for the slope.

This is to be compared to the full asymptotic expansion of (\ref{sim})
in $\ln(\xi)$ \cite{BO78}, the leading term corresponding to 
Voinov's solution:
\begin{equation}  
\label{exp}  
H'(\xi) = \left[3\delta\ln(\xi/\xi_0)\right]^{1/3}
\left\{1+\sum_{i=1}^{\infty}\frac{b_i}{(\ln(\xi/\xi_0))^i}\right\}.
\end{equation}  
All coefficients $b_1=1/3,b_2=-10/27,\dots$ are readily 
computable. Expanding (\ref{exp}) in $\delta$ and comparing to
(\ref{pert}) leads to the following structure of the expansion
of $\ln(\xi_0)$:
\begin{equation}  
\label{xi0}  
-3\ln(\xi_0) = \frac{1}{\delta} + \sum_{i=0}^{\infty}c_{i+1}\delta^i.
\end{equation}  
Substituting this back into (\ref{exp}), the large-$\xi$ behavior of the 
$H_i(\xi)$ in (\ref{pert}) is given in terms of the coefficients
$c_i$:
\begin{eqnarray}
\label{asymp}
\left.\begin{array}{l}
             H_1'(\xi) = \ln(\xi) + 1 + c_1/3, \\ 
 H_2'(\xi) = -\ln^2(\xi) - (2+2c_1/3)\ln(\xi)+ \\               
    + c_2/3-c_1^2/9-2c_1/3-10/3. 
                 \end{array}\right\}\xi\rightarrow\infty
\end{eqnarray}

To compute $c_1$ we have to solve (\ref{sim}) to first order
in $\delta$:
\begin{equation}  
\label{fo}  
H_1'''=-\frac{1}{\xi^2+\xi^{\alpha}}\equiv r(\xi),
\end{equation}  
with boundary conditions $H_1(0) = 0, H_1'(0)=1$, and $H_1''(\infty)=0$.
According to (\ref{asymp}), the constant $c_1$ is given by 
\begin{equation}  
\label{c1}  
1 + c_1/3 = \lim_{\xi\rightarrow\infty} H_1'(\xi) - \ln(\xi+1).
\end{equation}  
Integrating (\ref{fo}) twice, we can thus write 
\begin{eqnarray*}
1 + c_1/3 = \int_0^{\infty}\int_{\infty}^{\tilde{\xi}}
\left[r(\bar{\xi})+\frac{1}{(\bar{\xi}+1)^2}\right]d\bar{\xi}d\tilde{\xi}= \\
-\int_0^{\infty}\int_0^{\bar{\xi}}
\left[r(\bar{\xi})+\frac{1}{(\bar{\xi}+1)^2}\right]d\tilde{\xi}d\bar{\xi}= \\
\int_0^{\infty}
\left[\frac{\xi^{1-\alpha}}{\xi^{2-\alpha}+1}-
\frac{\xi}{(\xi+1)^2}\right]d\xi= \\
\left[\frac{1}{2-\alpha}\ln(\xi^{2-\alpha}+1)-
\ln(1+\xi)-\frac{1}{1+\xi}\right]^{\infty}_0 = 1\\
\end{eqnarray*}
{\it independent} of $\alpha$. Remarkably, $c_1$, the first-order correction
to $\xi_0$, always vanishes, regardless of how the length scale 
$\lambda$ is introduced near the contact line. 

The problem at second order can be tackled in precisely the same manner,
using the equation for the second-order problem 
\begin{equation}  
\label{so}  
H_2'''=H_1(\xi)\frac{2\xi+1}{\xi^2(\xi+1)^2},
\end{equation}  
where we have specialized to the standard case $\alpha=1$ for simplicity. 
The case $\alpha=0$ can be treated in exactly the same manner.
Integrating the first order equation (\ref{fo}) thrice gives 
\[
H_1 = [\ln(\xi+1)(\xi+1)^2 - \xi - \xi^2\ln(\xi)]/2,
\]
thus specifying the r.h.s. of equation (\ref{so}). The trick used to calculate 
$c_1$ at first order can be repeated at the next order, using the second
equation of (\ref{asymp}) and $c_1=0$. Simplifying the resulting 
double integral as before, we find
\begin{eqnarray*}
c_2/3 - 10/3 = \lim_{\xi\rightarrow\infty} H_2'(\xi) + \ln^2(\xi+1)
+ 2\ln(\xi+1) = \\
\int_0^{\infty} \left[\frac{-\xi H_1(\xi)(2\xi+1)+2\xi^3\ln(\xi+1)}
{\xi^2(\xi+1)^2}\right]d\xi  = \pi^2/6-7/2.
\end{eqnarray*}
In summary, we thus have $c_2 = (\pi^2-1)/2$ for $\alpha=1$ and 
$c_2 = (3\pi^2-4)/8$ for $\alpha=0$, which follows from a very 
similar calculation. 
It is evident that the same procedure can be repeated at arbitrary
order, although the calculation rapidly becomes analytically intractable. 

Rewriting the rescaled solution $H(\xi)$ in terms of the physical 
profile $h(x)$, we find the slope of the interface to be 
\begin{equation}  
\label{voinov}  
h'^3(x) - \theta_e^3 = 9Ca\ln(x/L)
\end{equation}  
for any capillary number. This is the form originally proposed by 
Voinov \cite{V76}, using more qualitative 
arguments. The length $L$ appearing inside the logarithm can be written 
as an expansion in the capillary number:
\begin{equation}  
\label{L}  
L = \frac{3^{1/(2-\alpha)}\lambda}{\theta_e}
\left[1 - \frac{3c_2 Ca}{\theta_e^3} + O(Ca^2)\right].
\end{equation}  
\begin{figure}
\includegraphics[width=1\hsize]{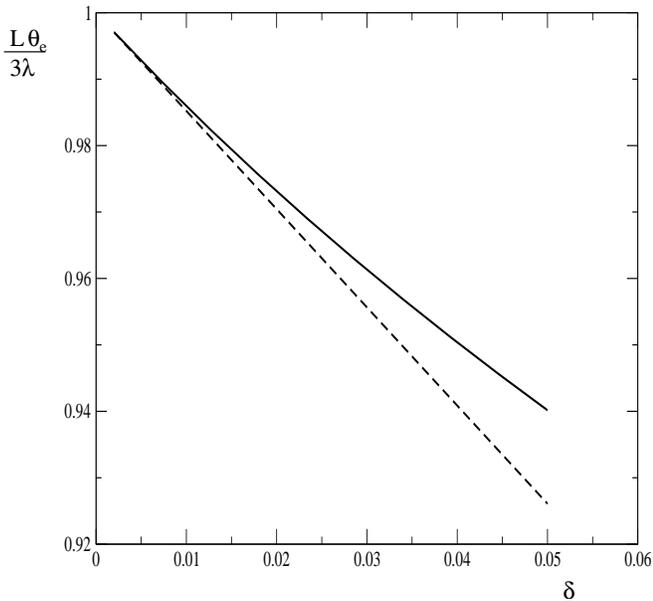}
\caption{\label{fig2} 
A comparison of (\ref{L}) with the numerical result for 
the characteristic length L, using numerical integration of (\ref{lub})
for $\alpha=1$. 
   }
\end{figure}

Integrating equation (\ref{lub}) numerically, a comparison with
(\ref{voinov}) can be made, giving $L$ as function of capillary 
number. In Fig.\ref{fig2} it is clearly seen that our expansion 
(\ref{L}) describes the initial departure from the leading
order result quite well. However, when the corrections amount to 
about $10\%$ of the leading order, higher order terms become 
important. Thus as a rough estimate, the present approach can 
be trusted if $\delta = Ca/\delta_e^3 \mbox{\ \raisebox{-.9ex}
{$\stackrel{\textstyle<}{\sim}$}\ } 0.05$. It would be interesting
to systematically investigate the dependence of $L$ on the capillary
number beyond that value. 

Comparing (\ref{voinov}) to de Gennes' result \cite{G86}
\begin{equation}  
\label{deGennes}  
(h'^2(x) - \theta_e^2)h'(x) = 6 Ca \ln(x/\lambda),
\end{equation}  
one finds that the two laws agree if the departure of
$h'(x)$ from $\theta_e$ is small. Beyond the leading 
order expansion of $h'(x)$ in $Ca$, however, (\ref{voinov})
and (\ref{deGennes}) are inconsistent. This casts doubts 
on the original argument for (\ref{deGennes}), which should also apply 
to the class of simple slip models considered here, and which 
has already been reviewed in a critical light in \cite{GHL90}.
In particular, this calls into question de Gennes' theory
\cite{G86} of contact line instability, which crucially uses 
(\ref{deGennes}).

Finally, it should be realized that we have assumed that the local 
profile as described by (\ref{lub}) can be treated independently. 
The only effect of the outer profile comes in with the boundary 
condition of vanishing curvature (\ref{bc}). Earlier work
\cite{H83,C86} includes the matching to an outer profile, 
but arrives at the same results as we do to leading order in 
$Ca$. At elevated capillary numbers, however, different parts of 
the solution may interact in a non-trivial way, and matching 
may be necessary. 

\acknowledgments
Thanks are due to Howard Stone for many discussions on the 
subject of contact lines, and to Rich Kerswell for help with
the analytical calculation of integrals.

\end{document}